\input{aipcheck}

\documentclass[
    ,final            % use final for the camera ready runs
  ]
  {aipproc}

\layoutstyle{6x9}

\newcommand{\BS}{Bethe--Salpeter }

\newcommand{\half}{\frac{1}{2}}

\newcommand{\s}{\!\cdot\!}

\newcommand{\ov}[1]{\overline{#1}}
\def\slr#1{\setbox0=\hbox{$#1$}           % set a box for #1
   \dimen0=\wd0                                 % and get its size
   \setbox1=\hbox{/} \dimen1=\wd1               % get size of /
   \ifdim\dimen0>\dimen1                        % #1 is bigger
      \rlap{\hbox to \dimen0{\hfil/\hfil}}      % so center / in box
      #1                                        % and print #1
   \else                                        % / is bigger
      \rlap{\hbox to \dimen1{\hfil$#1$\hfil}}   % so center #1
      /                                         % and print /
   \fi}

\begin{document}

\title{A Bethe--Salpeter Description of Light Mesons\footnote{Talk
presented by HW at the international {\it Scalar Meson Workshop}, 
Utica, NY, May 2003.}}

\author{H. Weigel}{
  address={Institute for Theoretical Physics, T\"ubingen University\\
Auf der Morgenstelle 14, D--72076 T\"ubingen, Germany}
}

\author{R. Alkofer}{
  address={Institute for Theoretical Physics, T\"ubingen University\\
Auf der Morgenstelle 14, D--72076 T\"ubingen, Germany}
}

\author{P. Watson}{
  address={Institute for Theoretical Physics I, Gie{\ss}en University\\
Heinrich--Buff--Ring 16, D--35392 Gie{\ss}en, Germany}
}

\begin{abstract}
We present a covariant approach to describe the low--lying scalar,
pseudoscalar, vector and axialvector mesons as quark--antiquark bound states.
This approach is based  on an effective interaction modeling of the
non--perturbative structure  of the gluon propagator that enters the 
quark Schwinger--Dyson and meson Bethe--Salpeter equations. 
We extract the meson  masses and compute the pion and kaon decay constants.
We obtain a quantitatively correct description for pions, kaons and vector
mesons while the calculated spectra of scalar and axialvector mesons
suggest that their structure is more complex than being
quark--antiquark bound states.
\end{abstract}

\maketitle

%%%%%%%%%%%%%%%%%%%%%%%%%%%%%%%%%%%%%%%%%%%%
%% MAINMATTER
%%%%%%%%%%%%%%%%%%%%%%%%%%%%%%%%%%%%%%%%%%%%

\section{Introduction}

Recently, the scalar mesons have attracted a lot of interest as the 
reanalysis of the pseudoscalar meson scattering data indicated the 
existence of a flavor SU(3) nonet in this channel~\cite{Syr1}. 
It is therefore desirable to gain deeper understanding of the constituent 
structure of the scalar mesons together with a comprehensive description
of the meson states in the other spin--parity channels. The ultimate 
goal would be to understand all low--lying meson states and resonances 
as non--perturbative bound states in Quantum Chromo Dynamics~(QCD). 

A relativistic framework for analyzing mesons as composite objects is  provided
by the Bethe--Salpeter equations that extract poles in the  quark--antiquark
scattering kernel\cite{alkofer00}. The attraction  needed to bind quarks and
antiquarks emerges from dressed multiple gluon  exchange. Thus the essential
ingredients to these equations are the quark and gluon propagators as well as
the quark--gluon vertex. In addition, these $n$-point Green's functions are
related by their  Schwinger--Dyson equations which are part of an infinite
tower of non--linear integral equations. There has been some progress in the
understanding of the infrared behavior of the gluon propagator from recent
Yang--Mills lattice measurements \cite{Mandula:1999nj} as well as from 
studies of the coupled system of gluon and ghost Schwinger--Dyson equations 
\cite{alkofer00,vonSmekal:1997is}. More recently, the coupled system of
gluon, ghost and quark propagators functions has been studied within
certain truncation schemes of the Schwinger--Dyson equations and
solutions have been obtained for various ans\"atze for the 
ghost--gluon vertices \cite{Fischer:2003rp}.  
Nevertheless, for phenomenological
applications the frequently adopted strategy is to model the gluon 
propagator as well as the quark--gluon vertex and consistently derive 
the quark propagator from its Schwinger--Dyson equation.

These types of calculations have a long history, for reviews see
refs.~\cite{alkofer00,roberts00}. Early versions adopted pointlike
gluon propagators in coordinate space that eventually lead to
Nambu--Jona--Lasinio (NJL) type models~\cite{nambu61,ebert86}, 
pointlike propagators in momentum space were also considered~\cite{Ja93}.
These models are particularly simple because either solving the
Schwinger--Dyson equation yields a free quark propagator or the
Bethe--Salpeter integral equations reduce to algebraic equations. 
The main target particularly of the NJL--model studies have
been the pseudoscalar mesons. It turned out that they can be
adequately described once the important feature of dynamical
chiral symmetry breaking is incorporated, {\it i.e.} the interaction 
is strong enough so that the resulting quark propagator develops
a non--zero constituent quark mass. Then the pseudoscalar mesons can 
be understood as the {\it would--be} Goldstone bosons of chiral 
symmetry breaking. However, these NJL--type models do not reflect
the confinement property of QCD and thus
binding can only be achieved kinematically, {\it i.e.} meson 
states with masses larger than twice the constituent quark mass 
cannot be described consistently. For that reason, model gluon 
propagators have been developed that yield quark propagators without 
poles for real momenta as an attempt to include the confinement 
phenomena~\cite{maris98,maris99}. Again these studies focused on 
pseudoscalar mesons~\cite{Ro96,Ma97} while a comprehensive investigation 
for the scalar, pseudoscalar, vector and axialvector mesons has not 
been carried out so far. Other studies~\cite{maris98,maris99}
made contact with perturbative QCD by considering a model gluon 
propagator that matches the pertinent anomalous dimension. This 
contribution has negligible effect on the meson properties, but its 
inclusion makes cumbersome the extraction of the solutions to the 
Schwinger--Dyson equations for the large time--like momenta that enter 
the Bethe--Salpeter equations. Such large time--like momenta need to 
be considered for mesons other than the pseudoscalars. Although this
is interesting we regard it an unnecessary technical complication
because we do not want to compute properties of mesons revealed only
at high momentum transfer. Rather we want to establish a model 
as simple as possible that we 
consider a pertinent starting point to study the structure and properties 
of low--lying mesons in a fully relativistic framework. Our model 
interaction is parameterized in form of a non--trivial gluon 
propagator that contains sufficient strength to cause dynamical 
chiral symmetry breaking.  For technical reasons it
turns out that a Gau{\ss}ian shape function for the propagator in
momentum space is most suitable. Essentially we consider this model 
propagator as an effective interaction that relativistically describes 
the binding of quarks and antiquarks to mesons. Furthermore, we take 
the quark--gluon vertex function to be the tree level one since this 
procedure provides a framework that is consistent with chiral symmetry 
when the ladder approximation for the Bethe--Salpeter equation is 
employed \cite{alkofer00,roberts00}. For approaches going beyond
ladder approximation see e.g.~ref.~\cite{Be96}. 

This talk, which is mainly based on ref.~\cite{Alkofer:2002bp},
is organized as follows: First we will introduce the
effective interaction and solve the Schwinger--Dyson equation for the
quarks. We will put particular emphasis on the analytic continuation
of the resulting quark propagator to time--like momenta that enter 
the Bethe--Salpeter equations. We will discuss the structure of the
Bethe--Salpeter equations and then present solutions. Finally 
we will conclude and suggest a possible extension of the current
approach in particular with regard to the possibility that the 
scalar meson might have to be considered as two--quark -- two--antiquark
bound states~\cite{Syr1,Ja77}.

%%%%%%%%%%%%%%%%%%%%%%%%%%%%%%%%%%%%%%%%%%%%%%%%%%%%%%%%%%%%%%%%%%
\section{The Quark Schwinger-Dyson Equation}

We take a Gau{\ss}ian form for dressing the model gluon propagator 
and write,
\begin{equation}
g^{2}G_{\mu\nu}^{ab}(q)=4\pi^2 D \delta^{ab}t_{\mu\nu}(q)
\frac{q^2}{\omega^2}\, \exp{\left(-\frac{q^{2}}{\omega^{2}}\right)}
\label{eq:gluon}
\end{equation}
where $\mu,\nu$ are Lorentz indices, $t_{\mu\nu}(q)$ is the transverse 
momentum projector and $a,b$ label color. While the coefficients in 
eq.~(\ref{eq:gluon}) are chosen to make subsequent equations more 
concise,~$D$ and~$\omega$ are dimensionful parameters that we will determine 
from fitting empirical data. The coefficient~$D$ sets the strength of the 
interaction and~$\omega$ is the value at which the scalar function in 
the parameterization is maximal. Hence~$\omega$ sets the interaction scale.  
The dressed gluon propagator~(\ref{eq:gluon}) is supposed to represent 
a sensible hadron model and hence one can envisage that~$\omega$ will have 
a value of several hundred~${\rm MeV}$. 

We interpret the effective interaction~(\ref{eq:gluon}) as the propagator 
(in Landau gauge) of a gluon that gets absorbed and emitted by the 
quarks that eventually get bound to form mesons. To completely define 
the interaction, we need to parameterize the quark--gluon coupling. 
To establish chiral symmetry we apply the rainbow--ladder approximation 
to the system of Schwinger--Dyson and Bethe--Salpeter equations. 
Then the quark--gluon coupling is given by the tree level
interaction vertex, $ig\gamma_\mu \frac{\lambda^a}{2}$, where 
$\lambda^a$ is a Gell--Mann matrix acting in color space. Note, that 
we have already included the coupling constant~$g$ in the definition of 
the effective interaction~(\ref{eq:gluon}).

The Schwinger--Dyson equation for the (inverse) quark propagator 
becomes 
\begin{equation}
S^{-1}(p)=i\slr{p}+m_0+\int\frac{d^4k}{(2\pi)^4}\,
\gamma_{\mu} S(k)\gamma_{\nu}\, g^2\, 
\frac{\lambda^a}{2}\frac{\lambda^b}{2}\,G_{\mu\nu}^{ab}(k-p)
\label{eq:qdse}
\end{equation}
where $m_0$ is the current mass of the considered quark. This
contribution represents the only explicit distinction between quarks 
of different flavors. Of course, its effects will implicitly propagate 
through the whole calculation. However, for notational simplicity
we will continue to suppress flavor labels. A suitable parameterization 
of the quark propagator is inspired by the form of a free fermion propagator
\begin{equation}
S(p)=\left[\frac{1}{i\slr{p}A(p^2)+B(p^2)}\right]\,.
\end{equation}
In solving the Schwinger--Dyson equation~(\ref{eq:qdse}) we have to 
find the scalar functions $A(p^2)$ and $B(p^2)$. It is also very 
instructive to define a mass function via $M(p^2)=B(p^2)/A(p^2)$. In 
particular $M(p^2=0)$ plays the role of a constituent quark mass and a
large value ($\gg m_0$) thereof signals dynamical chiral symmetry 
breaking.
 
We work in Euclidean space with Hermitian Dirac matrices
that obey $\{\gamma_{\mu},\gamma_{\nu}\}=2\delta_{\mu\nu}$ and
$\gamma_5=-\gamma_1\gamma_2\gamma_3\gamma_4$.  
Inserting the effective interaction~(\ref{eq:gluon}) and performing
the standard trace algebra, we then deduce the following coupled 
equations for the propagator functions 
\begin{eqnarray}
A(x) &=& 1+D\int_{0}^{\infty}\!\frac{dy\,y\,A(y)}{(yA^2(y)+B^2(y))}
\exp{\left\{-\frac{x+y}{\omega^{2}}\right\}}
\nonumber\\* && \hspace{2cm}\times
\left\{\left(1+\frac{y}{x}+2\frac{\omega^2}{x}\right)
I_2\left(\frac{2\sqrt{xy}}{\omega^2}\right)-2\frac{\sqrt{y}}{\sqrt{x}}\,
I_1\left(\frac{2\sqrt{xy}}{\omega^2}\right)\right\},
\nonumber\\
B(x)&=& m_0+D\int_{0}^{\infty}\!\frac{dy\,y\,B(y)}{(yA^2(y)+B^2(y))}
\exp{\left\{-\frac{x+y}{\omega^{2}}\right\}}
\nonumber \\ && \hspace{2cm}\times
\left\{\left(\frac{\sqrt{y}}{\sqrt{x}}
+\frac{\sqrt{x}}{\sqrt{y}}\right)I_1
\left(\frac{2\sqrt{xy}}{\omega^2}\right)-2I_2
\left(\frac{2\sqrt{xy}}{\omega^2}\right)\right\}\, ,
\label{eq:qdses}
\end{eqnarray}
where the four dimensional integral measure has been expanded such 
that $x=p^2$ and $y=k^2$. Furthermore $I_n$ are modified Bessel 
functions. Note, that as a particular feature of the
Gau{\ss}ian dressing function~(\ref{eq:gluon}) it has been
feasible to compute the angular integrals analytically. 

In a first step we solve eqs.~(\ref{eq:qdses}) for spacelike 
momenta, {\it i.e.} for real positive~$x$. Then we observe 
that the integrals on the $RHS$ of these equations only involve the 
propagator functions at real arguments $y$ and we can use them to 
numerically compute the propagator functions for {\it arbitrary 
complex} $x$.  At first sight, it appears that $A(x)$ and
$B(x)$ could not be consistently continued because the cut
along the negative $x$--axis (associated with $\sqrt{x}$) 
would yield different results when continuing in the 
upper or the lower half--plane and it would be impossible to 
resolve the ambiguity in $\sqrt{x}\to\pm i\sqrt{\xi}$ when 
continuing $x\to -\xi$. Fortunately, this is not an obstacle
because the modified Bessel functions $I_1$ and $I_2$ are 
respectively odd and even functions of their arguments. 
Thus we are free to choose either of the two signs above. 
For definiteness we work in the upper half--plane with 
$\sqrt{x}\to i\sqrt{\xi}$ along the negative half--line.

\begin{figure}
\centerline{
\includegraphics[height=.25\textheight]{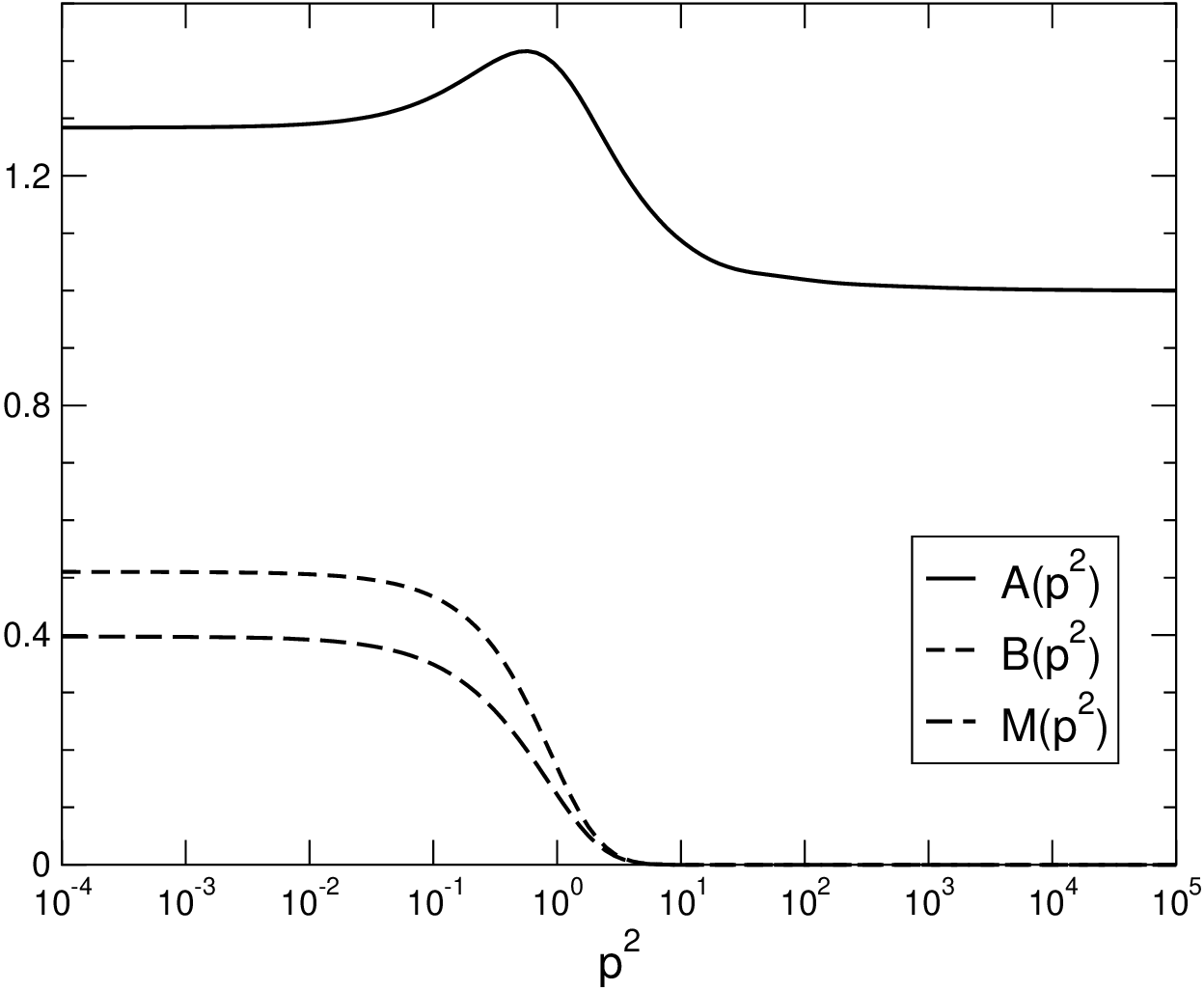}\hspace{1cm}
\includegraphics[height=.25\textheight]{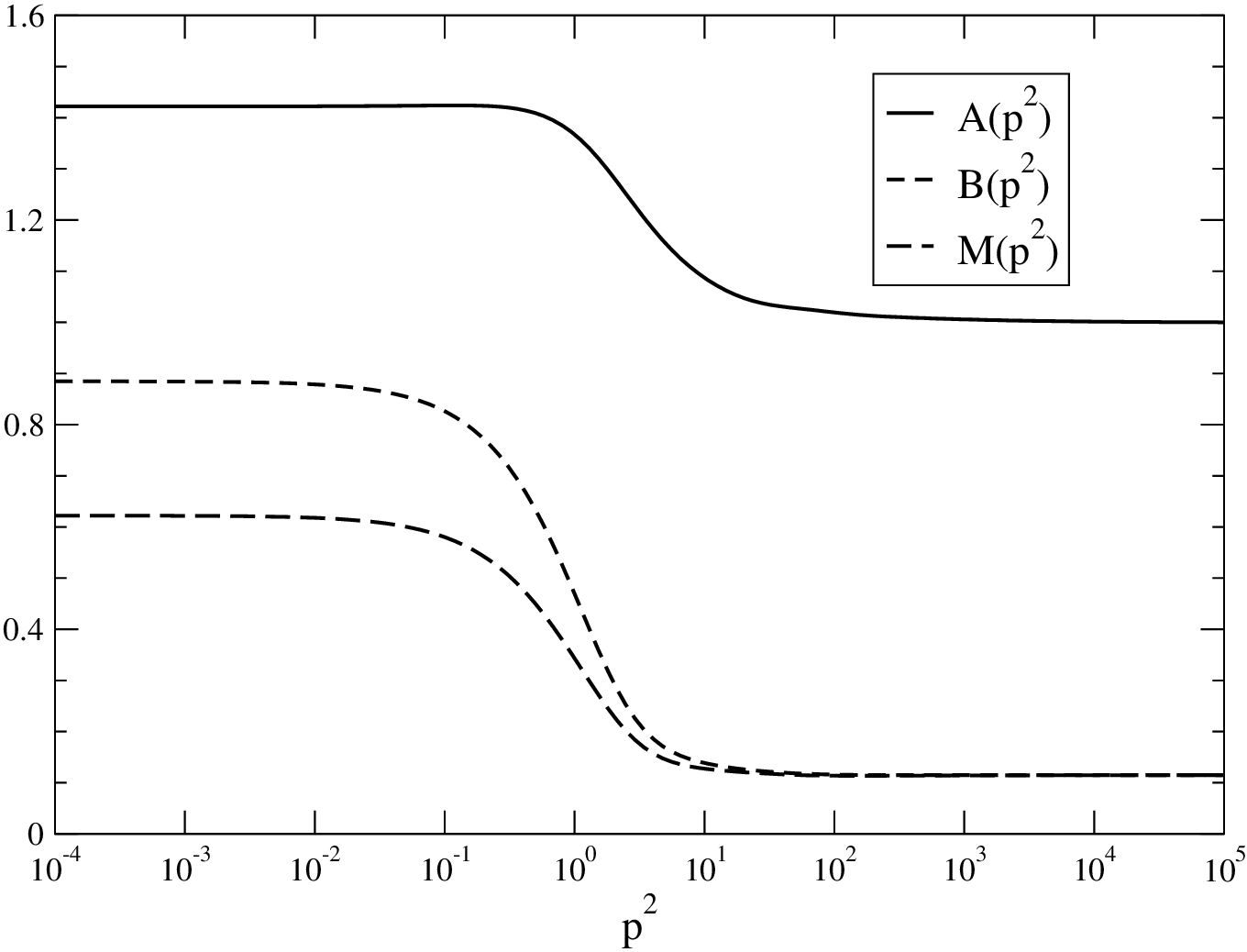}}
~\smallskip
\caption{Quark propagator functions (as a
function of the momentum squared).  The parameters are
$\omega=0.5$GeV, $D=16.0$GeV$^{-2}$. Left panel: $m_0=0$, right panel: $m_0=0.115$
GeV.
All units are ${\rm GeV}$.}
\label{fig:quark1}
\end{figure}

In Fig.\ \ref{fig:quark1} we show the quark propagator functions 
$A(p^2)$, $B(p^2)$ and $M(p^2)=B(p^2)/A(p^2)$ along the positive, real 
spacelike axis,  $p^2>0$.  
This is the numerical 
solution to the coupled equations (\ref{eq:qdses}), which we emphasize 
is the basis for the quark solutions for complex momenta.  The solution 
clearly shows that dynamical chiral symmetry breaking is occurring: the 
mass function $M(p^2)$ attains a sizable non--zero value, even in the
case that the bare quark mass $m_0$ is zero. This phenomenon 
is an example of genuinely non--perturbative behavior as dynamical 
mass generation cannot occur at any order in perturbation theory.  
Recall that the effective interaction (\ref{eq:gluon}) that enters the 
Schwinger--Dyson equations does not contain the perturbative UV behavior, 
rather it has an exponential damping at high momenta. This is manifested 
in the quark propagator functions as a sharp transition from the low 
momentum behavior to the bare values in the high momentum region. This 
transition occurs at about $1{\rm GeV}$.  

%%%%%%%%%%%%%%%%%%%%%%%%%%%%%%%%%%%%%%%%%%%%%%%%%%%%%%%%%%%%%%%%%%
\section{The Bethe-Salpeter Equation}

Having obtained the quark propagators in the complex plane from the 
Schwinger--Dyson equations we have collected all ingredients for the 
Bethe--Salpeter integral equations. They will ultimately yield the 
quark meson vertex functions, $\Gamma$, that describe mesons as 
bound quark--antiquark pairs. These vertex functions for bound
states are determined from the condition that the four quark
scattering amplitude develops a pole at $P^2=-M^2$, where 
$M$ is the mass of the bound state meson.

The vertex functions resulting from the Bethe--Salpeter equation
are characterized by three momenta out
of which only two are linearly independent due to momentum conservation
at the vertex. If we denote the meson momentum $P$ and the momentum
of the incoming quark $p+\xi P$ then the momentum of the outgoing
quark (= incoming antiquark) is $p+(\xi-1)P$. This suggests to label
the vertex functions by $p$ and $P$: $\Gamma(p,P)$. We have 
introduced the arbitrary momentum partition parameter~$\xi\in[0,1]$. 
Due to strict relativistic covariance the results for physical
observables do not depend on $\xi$.  We have studied the 
$\xi$--dependence of our numerical results exhaustively in 
ref.~\cite{Alkofer:2002bp}. The confirmed $\xi$--independence 
represents an {\it a posteriori} 
validation for the relativistic covariance of our computations. 

We now turn to the main target of 
our studies, the Bethe--Salpeter integral equations for the vertex 
function~$\Gamma(p,P)$ in ladder 
approximation~\cite{alkofer00,roberts00,tandy}:
\begin{equation}
\Gamma(p;P)=-\frac{4}{3}\int\frac{d^4k}{(2\pi)^4}\, 
\left[\gamma_{\nu}\,S(k+\xi P)\,\Gamma(k;P)\,S(k+(\xi-1)P)\,
\gamma_{\mu}\right]\,g^2 G_{\mu\nu}(k-p)\,.
\label{eq:bse}
\end{equation}
Here we have factorized the color factors in the effective interaction, 
$G_{\mu\nu}^{ab}(q)=\delta^{ab}G_{\mu\nu}(q)$ and performed the 
corresponding trace. The flavor content of the meson is not made 
explicit in eq.~(\ref{eq:bse}) as we have suppressed the 
flavor labels in the quark propagators. It is understood 
that the two propagators in eq.~(\ref{eq:bse}) are taken 
such as to account for the flavor quantum numbers of the 
considered meson. In the model that we will consider, the up and 
down quarks will be assumed to have equal current masses ($m_0$ in 
eq.~(\ref{eq:qdses})) and thus also identical 
propagator functions $A(x)$ and $B(x)$. For the light quarks, which 
should give rise to the familiar $SU(3)_f$ nonet, we are thus left with 
three representatives of each of the multiplets that are distinguished 
by their isospin number, $I=0,\half,1$ . We must also specify the
meson angular momentum and parity. This is reflected by the 
Dirac and Lorentz decomposition of the meson vertex functions. This 
decomposition is known in the literature and
here we follow ref.~\cite{llew}. For the pseudoscalar 
channel~($J^{P}=0^{-}$) we take
\begin{equation}
\Gamma^{(P)}(p;P)=\gamma_{5}\left[\Gamma^{(P)}_0(p;P)
-i\slr{P}\Gamma^{(P)}_1(p;P)-i\slr{p}\Gamma^{(P)}_2(p;P)
-\left[\slr{P},\slr{p}\right]\Gamma^{(P)}_3(p;P)\right]\, .
\label{eq:pseu}
\end{equation}
The decomposition for a scalar ($J^{P}=0^{+}$) meson reads
\begin{equation}
\Gamma^{(S)}(p;P)=\Gamma^{(S)}_0(p;P)-i\slr{P}\Gamma^{(S)}_1(p;P)
-i\slr{p}\Gamma^{(S)}_2(p;P)
-\left[\slr{P},\slr{p}\right]\Gamma^{(S)}_3(p;P).
\label{eq:scal}
\end{equation}
The vector ($J^{P}=1^{-}$) channel involves eight scalar functions
\begin{eqnarray}
\Gamma^{(V)}_{\mu}(p;P)\hspace{-0.2cm} &=&\hspace{-0.2cm} 
\left[\gamma_{\mu}-\frac{P_{\mu}\slr{P}}{P^2}\right]\hspace{-0.1cm}
\left[i\Gamma^{(V)}_0(p;P)+\slr{P}\Gamma^{(V)}_{1}(p;P)
-\slr{p}\Gamma^{(V)}_2(p;P)+i\left[\slr{P},\slr{p}\right]
\Gamma^{(V)}_3(p;P)\right]\nonumber\\
&&+\left[p_{\mu}-\frac{P_{\mu}p\s P}{P^2}\right]
\left[\Gamma^{(V)}_{2}(p;P)+2i\slr{P}\Gamma^{(V)}_3(p;P)\right]
\label{eq:vect}  \\ &&
\hspace{-1.2cm}
+\left[p_{\mu}-\frac{P_{\mu}p\s P}{P^2}\right]\hspace{-0.1cm}
\left[\Gamma^{(V)}_4(p;P)+i\slr{P}\Gamma^{(V)}_5(p;P)
-i\slr{p}\Gamma^{(V)}_6(p;P)+\left[\slr{P},\slr{p}\right]
\Gamma^{(V)}_7(p;P)\right]\, .
\nonumber
\end{eqnarray}
In the axialvector channel we have two modes that are distinguished
by their charge conjugation properties~\cite{llew}. For $J^{PC}=1^{++}$ 
the decomposition is
\begin{eqnarray}
\Gamma^{(A)}_{\mu}(p;P) \hspace{-0.23cm} &=&\hspace{-0.23cm}
\gamma_{5}\hspace{-0.1cm}
\left[\gamma_{\mu}\hspace{-0.05cm}-\hspace{-0.05cm}
\frac{P_{\mu}\slr{P}}{P^2}\right]
\hspace{-0.1cm}
\left[i\Gamma^{(A)}_0(p;P)+\slr{P}\Gamma^{(A)}_{1}(p;P)
-\slr{p}\Gamma^{(A)}_2(p;P)+i\left[\slr{P},\slr{p}\right]
\Gamma^{(A)}_3(p;P)\right]
\nonumber\\ &&
+\gamma_{5}\left[p_{\mu}-\frac{P_{\mu}p\s P}{P^2}\right]
\left[\Gamma^{(A)}_{2}(p;P)+2i\slr{P}\Gamma^{(A)}_3(p;P) \right]\, ,
\label{eq:axip}
\end{eqnarray}
while the $J^{PC}=1^{+-}$ mode is decomposed as 
\begin{eqnarray}
\Gamma^{(\tilde{A})}_{\mu}(p;P) \hspace{-0.23cm} &=&\hspace{-0.23cm}  
\gamma_{5}\left[p_{\mu}\hspace{-0.05cm}-\hspace{-0.05cm}
\frac{P_{\mu}p\s P}{P^2}\right]\hspace{-0.1cm}
\left[\Gamma^{(\tilde{A})}_1(p;P)+i\slr{P}\Gamma^{(\tilde{A})}_2(p;P)
\right. \nonumber \\ && \hspace{3cm} \left.
-i\slr{p}\Gamma^{(\tilde{A})}_3(p;P)+\left[\slr{P},\slr{p}\right]
\Gamma^{(\tilde{A})}_4(p;P)\right].
\label{eq:axim}
\end{eqnarray}
In what follows we will omit the superscripts that label the 
spin and parity channels because these channels do not mix 
and there should hence be no confusion.

The solution to the Bethe--Salpeter equation not only yields the
meson masses but also the meson quark vertex functions that
can be used to compute meson properties. Here we will focus
on the pseudoscalar decay constants $f_{\pi}$ and $f_K$. In order 
to calculate these, we first have to normalize the vertex functions 
$\Gamma(p;P)$.  The \BS equation is a homogeneous equation, and 
thus needs an additional normalization condition. As mentioned 
previously, that condition is obtained from demanding the pole in 
the four--quark Green's function to be unity. For equal momentum 
partitioning, ({\it i.e.} for $\xi=1/2$ only) it reads~\cite{tandy}
\begin{eqnarray}
2P_{\mu}&=&3\int\frac{d^4k}{(2\pi)^4}\, 
{\rm Tr}\left\{
\ov{\Gamma}(k,-P)\frac{\partial S(k+P/2)}{\partial P_{\mu}}
\Gamma(k,P)S(k-P/2)
\right.\nonumber\\* && \hspace{3cm}\left.
+\ov{\Gamma}(k,-P)S(k+P/2)\Gamma(k,P)
\frac{\partial S(k-P/2)}{\partial P_{\mu}}\right\}
\label{norm}
\end{eqnarray}
where the trace is over Dirac matrices.  The conjugate 
vertex function $\ov{\Gamma}$ is defined as
\begin{equation}
\ov{\Gamma}(p,-P)= C\Gamma^T(-p,-P)C^{-1}\, ,
\end{equation}
where $C=-\gamma_2\gamma_4$ is the charge conjugation matrix. 
The quark propagator derivatives are 
calculated by differentiating the quark Schwinger--Dyson
equations~(\ref{eq:qdses}) analytically and 
then numerically integrating the corresponding expressions.

The decay constants are finally obtained from the coupling of
the axial current to the quark loop~\cite{tandy}
\begin{equation}
f=\frac{3}{M^2}\int\frac{d^4k}{(2\pi)^4}\, {\rm Tr}
\left\{\Gamma(k,-P)S(k+P/2)\gamma_5\slr{P}S(k-P/2)\right\}\,.
\end{equation}

The primary subject of this talk is to extract the bound state masses
for the various flavor combinations and angular momentum channels.
The corresponding projection results in sets of coupled 
equations for the $\Gamma_i$. After carrying out
two of the three angular integrals analytically we are left with 
functions of the squared momenta $p^2$ and $P^2$ as well as the
angle between $p$ and $P$: $z=p\s P/\sqrt{p^2P^2}$. The $z$--dependence
is analyzed by an expansion in Chebyshev polynomials $T_k$
\begin{equation}
\Gamma_i(p;P)=\sum_k (i)^k \Gamma_i^k(p^2;P^2)T_k(z)\, .
\label{eq:cheb}
\end{equation}
Since the $T_k$ form an orthonormal set, we can project the equations 
for the Dirac components onto $\Gamma_i^k(p^2;P^2)$. Finally, the 
$k^2$--integral in the Bethe--Salpeter equation~(\ref{eq:bse}) is 
implemented numerically as a matrix equation for the unknown 
$\Gamma_i^k(p_j^2;P^2)$, $p_j^2$ being the discrete values of the 
momentum squared. The kernel, $K$ of that matrix parametrically depends 
on the meson momentum $P^2$. We solve that matrix equation as an 
eigenvalue problem by tuning the meson momentum to $P^2=-M^2$, such
that ${\rm Det}(1-K)=0$. This yields the desired meson mass~$M$.

In the discussion of numerical results we note that there 
are four model parameters, 
$\omega,D,m_u$ and $m_s$ that we first have to fit to empirical data.  To 
this end, we initially choose the pseudoscalar meson observables 
$M_{\pi},M_{K}$ and $f_{\pi}$.  Then one parameter remains unconstrained 
by the pseudoscalar sector alone. However, the condition that the quark 
propagator function reflects dynamical chiral symmetry breaking, 
{\it i.e.} that $M(p^2=0)\approx 0.5{\rm GeV}$ leaves only a small window 
for the remaining choice.
All other masses and decay constants are subsequently model predictions.
The resulting model parameter and the predicted kaon decay constant $f_K$,
that is unexpectedly well reproduced, are shown in Table~\ref{tab:res1}. 
\begin{table}
\begin{tabular}{c|c|c|c||c|c|c|c}
$\omega$ & $D$ & $m_u$ & $m_s$ & $M_{\pi}$ 
& $f_{\pi}$ & $M_K$ & $f_K$ \\ \hline
0.40 & 45.0 & $5\times10^{-3}$ & $0.120$ 
& 0.135 & 0.131 & 0.496 & 0.164 \\ \hline
0.45 & 25.0 & $5\times10^{-3}$ & $0.120$ 
& 0.135 & 0.131 & 0.496 & 0.163 \\ \hline
0.50 & 16.0 & $5\times10^{-3}$ & $0.115$ 
& 0.137 & 0.133 & 0.492 & 0.164 \\ \hline
\multicolumn{3}{c}{experiment \cite{pdg}}& 
& 0.135 & 0.131 & 0.498 & 0.160
\end{tabular}
\caption{Parameter sets used and fit results for the
pseudoscalar mesons.  $M_{\pi}$, $f_{\pi}$ and $M_K$ are used
as input, $f_K$ is predicted. All units are ${\rm GeV}$.}
\label{tab:res1}
\end{table}
The subsequently predicted meson masses are shown in 
Tables~\ref{tab:res2}-\ref{tab:res5}. In all cases we have
the inequalities $M_{u\bar{u}}<M_{u\bar{s}}<M_{s\bar{s}}$,
where the subscript labels the flavor content. These relations
just reflect the quark--antiquark picture that is implicit
in the present Bethe--Salpeter approach.

Obviously both the pseudoscalar (table~\ref{tab:res1}) and vector 
mesons (table~\ref{tab:res2}) can be very well described within our 
model with the choice $\omega\approx 0.5{\rm GeV}$. Our results agree with 
a previous analysis of the vector mesons based on an effective
interaction which included the perturbative type term~\cite{maris99}.
This shows that such terms do not have a large effect on the meson
masses, at least for the pseudoscalar and vector cases. Indeed, in
the context of low--energy meson phenomenology we conclude that 
the logarithmic tail, and its associated renormalization represent
an unnecessary obfuscation.
\begin{table}
\begin{tabular}{c|c|c|c||c|c|c}
$\omega$ & $D$ & $m_u$ & $m_s$ & $M_{\rho}$ & 
$M_{K^*}$ & $M_{\phi}$ \\ \hline
0.40 & 45.0 & $5\times10^{-3}$ & $0.120$  & 0.748 & 0.939 & 1.072 \\ \hline
0.45 & 25.0 & $5\times10^{-3}$ & $0.120$ & 0.746 & 0.936 & 1.070 \\ \hline
0.50 & 16.0 & $5\times10^{-3}$ & $0.115$ & 0.758 & 0.946 & 1.078 \\ \hline
\multicolumn{3}{c}{experiment \cite{pdg}}&
& 0.770 & 0.892 & 1.020
\end{tabular}
\caption{Results for the vector mesons. All units are ${\rm GeV}$.}
\label{tab:res2}
\end{table}

The situation for the scalar mesons (table~\ref{tab:res3}) is not quite
that clear. To begin with the particle data group~\cite{pdg} does not 
provide a clear picture in this channel but only quotes a wide range for 
the mass of the lowest scalar (0.4 -- 1.2 GeV). More detailed studies
of the pseudoscalar scattering amplitudes revealed that the assignment 
of the scalar meson nonet is not at all established~\cite{Syr1}. 
In particular, these mesons may not be simple quark--antiquark bound
states but {\it e.g.} might contain sizable admixture of 
2quark--2antiquark pairs~\cite{Ja77}. In that respect we might interpret
our results as a quark--antiquark model prediction for scalar mesons.
Our results suggest that such a picture is too simple for these mesons.
One might also speculate that the adopted ladder approximation could 
be insufficient.
\begin{table}
\begin{tabular}{c|c|c|c||c|c|c}
$\omega$ & $D$ & $m_u$ & $m_s$ & $M_{u\ov{u}}$ 
& $M_{u\ov{s}}$ & $M_{s\ov{s}}$ \\ \hline
0.40 & 45.0 & $5\times10^{-3}$ & $0.120$  & 0.700 & 0.917 & 1.096 \\ \hline
0.45 & 25.0 & $5\times10^{-3}$ & $0.120$ & 0.675 & 0.908 & 1.099 \\ \hline
0.50 & 16.0 & $5\times10^{-3}$ & $0.115$ & 0.645 & 0.903 & 1.113
\end{tabular}
\caption{Results for the scalar mesons. The subscripts of $M$ 
denote the flavor content. All units are ${\rm GeV}$.}
\label{tab:res3}
\end{table}

For the axialvector mesons we have two channels that are 
distinguished by their charge conjugation properties, {\it cf.}
tables~\ref{tab:res4} and~\ref{tab:res5}. The quark--antiquark
pairs that are bound to axialvector modes with negative
charge conjugation eigenvalue tend to be lighter than those with 
the positive eigenvalue but otherwise equal quantum numbers. 
Generally we find that our predictions are lower than the 
assignments made by the particle data group~\cite{pdg}.
\begin{table}
\begin{tabular}{c|c|c|c||c|c|c}
$\omega$ & $D$ & $m_u$ & $m_s$ & $M_{u\ov{u}}$ 
& $M_{u\ov{s}}$ & $M_{s\ov{s}}$ \\ \hline
0.40 & 45.0 & $5\times10^{-3}$ & $0.120$  & 0.804 & 0.994 & 1.128 \\ \hline
0.45 & 25.0 & $5\times10^{-3}$ & $0.120$ & 0.858 & 1.047 & 1.182 \\ \hline
0.50 & 16.0 & $5\times10^{-3}$ & $0.115$ & 0.912 & 1.098 & 1.230 \\ \hline
\multicolumn{3}{c}{experiment \cite{pdg}}&                      
& 1.230 & 1.270 & 1.170 ?
\end{tabular}
\caption{Results for the axial-vector ($J^{PC}=1^{+-}$) mesons.  
The question mark indicates that the PDG did not assign the
charge conjugation property of the respective resonance.
All units are ${\rm GeV}$.}
\label{tab:res4}
\end{table}
\begin{table}
\begin{tabular}{c|c|c|c||c|c|c}
$\omega$ & $D$ & $m_u$ & $m_s$ & $M_{u\ov{u}}$ 
& $M_{u\ov{s}}$ & $M_{s\ov{s}}$ \\ \hline
0.40 & 45.0 & $5\times10^{-3}$ & $0.120$  & 0.917 & 1.117 & 1.253 \\ \hline
0.45 & 25.0 & $5\times10^{-3}$ & $0.120$ & 0.918 & 1.124 & 1.270 \\ \hline
0.50 & 16.0 & $5\times10^{-3}$ & $0.115$ & 0.927 & 1.140 & 1.292 \\ \hline
\multicolumn{3}{c}{experiment \cite{pdg}}&            
& 1.230 & 1.270 & 1.282
\end{tabular}
\caption{Results for the axial-vector ($J^{PC}=1^{++}$) mesons.  
All units are ${\rm GeV}$.}
\label{tab:res5}
\end{table}

We recognize from our results that the model predictions change 
only slightly within the large range of considered model parameters. 
This confirms that meson static properties are not too sensitive to the
conjectural parameter dependence of the timelike quark propagator 
functions. Presumably meson properties whose computation involves 
larger timelike momenta will exhibit a stronger sensitivity.

Already from table~\ref{tab:res4} we observe that by increasing
$\omega$ the predicted mass of the $J^{PC}=1^{+-}$ meson with pion 
flavor quantum numbers approaches the empirical mass.
We therefore further increased
$\omega$ according to the rules discussed above. For $\omega\sim0.8GeV$ we
reproduced the empirical value for the mass in that channel. However, this 
happened at the expense of significantly lowering $f_K$ and loosing
the proper description of the vector mesons. We recall that the parameter 
$\omega$ has a physical interpretation as the location of the maximum of 
the interaction. Thus $\omega=0.8GeV$ seems intuitively too large for 
low--energy hadron physics and an unsatisfactorily description of the 
$0^-$ and $1^-$ mesons comes without surprise.

For non--diagonal flavor structures such as $u\bar{s}$, charge 
conjugation actually is not a sensible quantum number and the 
corresponding axial vector mesons $1^{++}$ and $1^{+-}$ may mix.
In table~\ref{tab:res6} we present the results obtained from the full 
calculation that combines the Dirac decompositions~(\ref{eq:axip}) 
and~(\ref{eq:axim}).
\begin{table}[h]
\begin{tabular}{c|c|c|c||c|c|c}
$\omega$ & $D$ & $m_u$ & $m_s$ & $M_{u\ov{u}}$
& $M_{u\ov{s}}$ & $M_{s\ov{s}}$ \\ \hline
0.40 & 45.0 & $5\times10^{-3}$ & $0.120$ & 0.807 & 0.990 & 1.131 \\ \hline
0.45 & 25.0 & $5\times10^{-3}$ & $0.120$ & 0.861 & 1.040 & 1.185 \\ \hline
0.50 & 16.0 & $5\times10^{-3}$ & $0.115$ & 0.915 & 1.085 & 1.233 \\ 
\end{tabular}
\caption{Results for the axial-vector mesons allowing for mixing 
of the Dirac structures in eqs.~(\protect\ref{eq:axip}) 
and~(\protect\ref{eq:axim}). All units are ${\rm GeV}$.}
\label{tab:res6}
\end{table}
Since our Bethe--Salpeter formalism only yields the lowest 
mass eigenstate within a given channel, the results presented
in table~\ref{tab:res6} should be compared to those in 
table~\ref{tab:res4}. The tiny changes for the flavor 
diagonal mesons are numerical artifacts. Surprisingly the
changes for the non--diagonal flavor structure are also
only of the order 1\%. This suggests only a small mixing
between the $1^{++}$ and $1^{+-}$ states with the flavor
structure $u\bar{s}$ and the $1^{++}$ and $1^{+-}$ channels 
represent good approximations to the actual eigenstates.

\begin{figure}[t]
\centerline{
\includegraphics[height=.25\textheight,width=.5\textwidth]{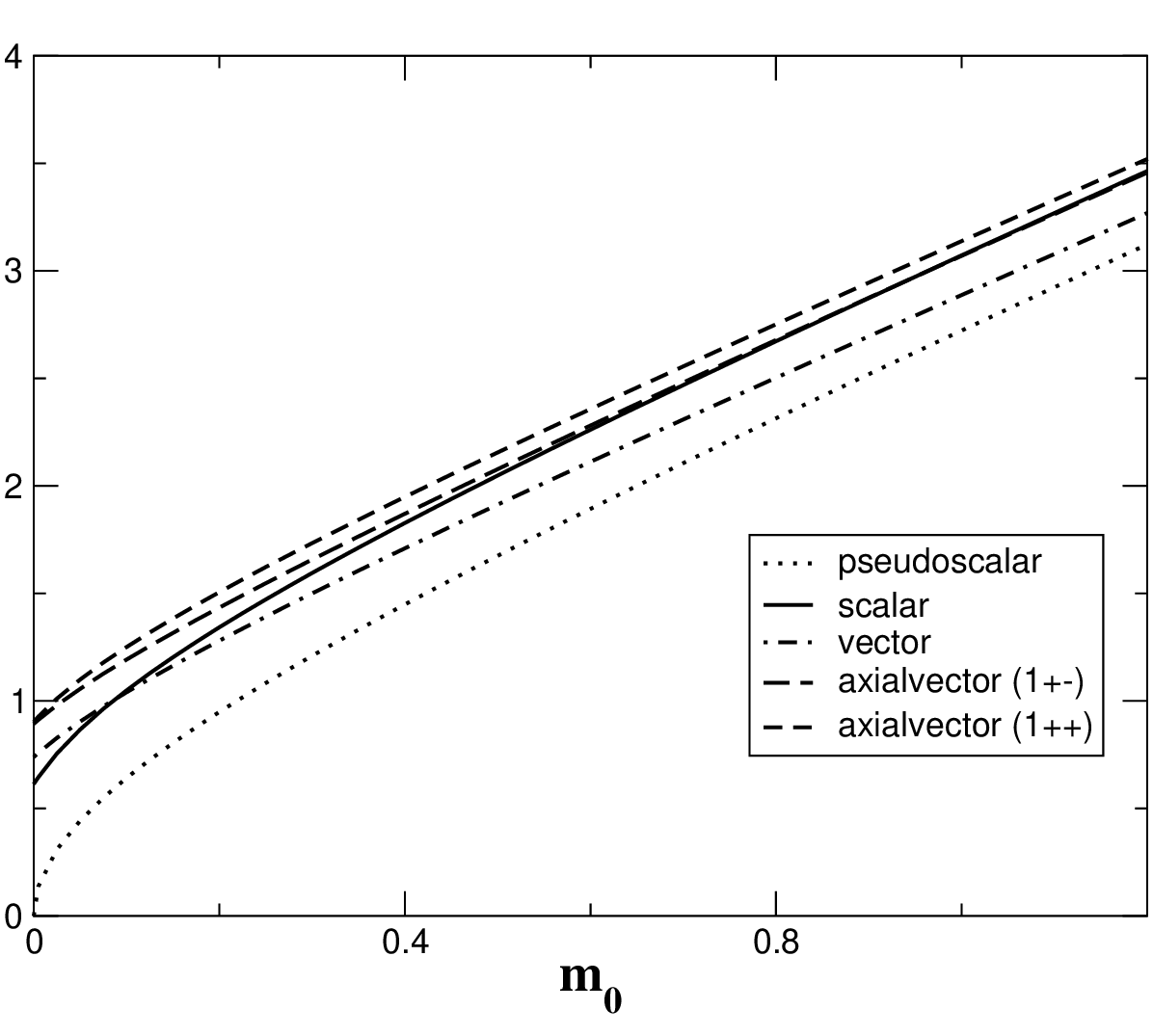}}
\caption{Meson masses as a function of the (equal) quark mass, 
$m_0$.  $\omega=0.5GeV, D=16GeV^{-2}$.  All units are ${\rm GeV}$.}
\label{fig:mass1}
\end{figure}

We can extend our model beyond the light flavors up, down 
and strange. The only modification is the increase of the current quark 
mass, $m_0$. In Fig.\ \ref{fig:mass1} we show how the meson 
masses increase as the (equal) quark masses are increased into the 
charm sector $m_c=1.125{\rm GeV}$.  
Exactly the same numerical code is used to construct these solutions to 
the Schwinger--Dyson and Bethe--Salpeter equations as for the light 
flavors. Clearly seen is the smooth way the masses increase from the 
chiral limit ($m_0=0$) into the heavy quark sector ($m_0=m_c$). This 
represents a convincing indicator for the stability of our technique. The 
$c\ov{c}$-meson masses can be loosely extracted (table~\ref{tab:res7}) 
and the data are surprisingly well reproduced. 
\begin{table}[t]
\begin{tabular}{l||c|c|c|c|c}
$J^{P(C)}$ & 
$0^-$ & $1^-$ & $0^+$ & $1^{++}$ & $1^{+-}$ \\ \hline
$M_{c\ov{c}}$ & 2.97 & 3.13 & 3.32 & 3.38 & 3.31\\ \hline
experiment \cite{pdg} &$\eta_c$: 2.98 & $J/\psi$: 3.10 &
$\chi_{c0}$: 3.42 & $\chi_{c1}$: 3.51 & ?
\end{tabular}
\caption{Predicted masses of $c\ov{c}$-meson states.  
$\omega=0.4,D=45.0$, $m_c=1.125$. 
$m_c$ is fitted approximately from the $\eta_c$ mass. \bigskip }
\label{tab:res7}
\end{table}
The lack of the correct UV behavior for the gluon is 
seemingly at odds with the scales present.  However, the present results 
suggest that the Bethe--Salpeter equation is capable of describing 
{\it all} the angular momentum states equally well in the charm quark sector.

%%%%%%%%%%%%%%%%%%%%%%%%%%%%%%%%%%%%%%%%%%%%%%%%%%%%%%%%%%%%%%%%%%
\section{Summary and Outlook}

In this talk we have presented a study of 
the low--lying mesons as quark--antiquark 
bound states in a covariant approach using an effective interaction.
This interaction is characterized by gluon exchange with the gluon 
propagator being dressed by a Gau{\ss}ian shape function. The interaction 
is completed by the quark--gluon vertex that we take to be the tree--level 
perturbative 
one. In this manner the rainbow--ladder approximation to the system
of Schwinger--Dyson and Bethe--Salpeter equation accounts for chiral 
symmetry. With this effective interaction, we have then consistently 
treated this system of integral equations by precisely implementing 
the quark propagator functions that solve the Schwinger--Dyson equations 
into the Bethe--Salpeter equations. Once the effective interaction
exceeds a certain strength, the Schwinger--Dyson equations exhibit 
dynamical chiral symmetry breaking and the pseudoscalar mesons emerge 
as {\it would--be} Goldstone bosons. We have then used observed properties 
of the pseudoscalar mesons to determine the model parameters. The kaon 
decay constant represents a model prediction. It turned out to be in 
good agreement with the empirical data. Furthermore our results for the 
vector meson masses match the experimental data. The situation in the 
scalar channel is less satisfying. As we solely consider the mesons 
as bound states of quark--antiquark pairs, it is not surprising that 
the mass eigenvalues increase with the strangeness content. On the other 
hand it is astonishing that for current quark masses, $m_0\ge0.2{\rm GeV}$, 
the lightest scalar mesons turn out to be heavier than the lightest 
vector mesons. When discussing these results it must be noted that the 
role and structure 
of the scalar mesons is still under intense debate. In particular, the 
question whether they should indeed be considered as quark--antiquark 
bound states is not yet completely resolved. There are 
indications, see e.g.\ ref.\ \cite{Syr1} and references therein,
that the scalar meson masses should 
actually decrease with the strangeness content of these mesons. This can
be understood if these mesons are considered as 2quark--2antiquark 
bound states in the sense of diquark--antidiquark systems~\cite{Ja77}. 

As an outlook we mention that there is an elegant way to extend the 
present model to incorporate such degrees of freedom. The Bethe--Salpeter 
treatment can be straightforwardly extended to study bound states of 
diquark--antidiquark pairs, once a binding mechanism is established. This 
could either be achieved by a gluon exchange similar to eq.~(\ref{eq:gluon})
or by quark exchange between a quark and a diquark. The latter
approach has been intensively studied and the corresponding
vertex is known from modeling baryon properties~\cite{tuediq}.
It will also be interesting to see whether these additional degrees of 
freedom will also affect the mass predictions for the axialvector mesons
that currently tend to be on the low side. Investigations in this 
direction are in progress.

%%%%%%%%%%%%%%%%%%%%%%%%%%%%%%%%%%%%%%%%%%%%%%%%
%% BACKMATTER
%%%%%%%%%%%%%%%%%%%%%%%%%%%%%%%%%%%%%%%%%%%%%%%%

\begin{theacknowledgments}
One of us (HW) would like to thank the organizers, in particular
A. Fariboz, for providing this worthwhile and pleasant workshop.

\noindent
This work has been supported by
DFG~(Al-297/3--3, Al-297/3--4, We-1254/3--2, We-1254/4--2)
and  COSY~(contract nos. 41376610, 41139452).
\end{theacknowledgments}

%\small \baselineskip =14pt
%%%%%%%%%%%%%%%%%%%%%%%%%%%%%%%%%%%%%%%%%%%%%%%%%%%%%%%%%%%%%%%%%%

\end{document}